\documentclass[pra,preprint,showpacs,showkeys,amsfonts]{revtex4}
\usepackage{graphicx}
 \RequirePackage{times}
\RequirePackage{mathptm}
\begin{document}

\def\ttt{{\rm Tr} }
\def\diag{{\rm diag} }
%\def\frak{\cal }
%\def\Bbb{\bf }
%\sloppy

\title{What could be more practical than a good interpretation?}
\author{Karl Svozil}
 \email{svozil@tuwien.ac.at}
\homepage{http://tph.tuwien.ac.at/~svozil}
\affiliation{Institut f\"ur Theoretische Physik, University of Technology Vienna,
Wiedner Hauptstra\ss e 8-10/136, A-1040 Vienna, Austria}

\begin{abstract}
Quantum mechanics is more than the derivation of
straightforward theorems about vector spaces,
Hilbert spaces and functional analysis.
In order to be applicable to experiment and technology,
those theorems need interpretation and meaning.
Interpretation is to the formalism
what a
scaffolding in architecture and
building construction is to the completed building.
\end{abstract}


\maketitle

Not long ago, we were sitting in
the university cafeteria with a gifted math student.
The student listened to the quantum mechanical problems
some condensed matter physicists were working on.
After a while he jumped up and shouted,
``but this is merely linear vector space theory!''
He seemed disappointed.
At least from a formal point of view, this exclamation appears not to be totally unreasonable.
I often have the chance to observe the astonishment of mathematicians,
for instance group theorists,
when I watch them browsing through venerable physics journals.
They point to numbered equations, lemmas, theorems and conjectures.
While shaking their heads disparagingly, they tell me that these results have been known for a long, long time,
So, why are mathematicians working in functional analysis
not the ``better'' quantum physicists?

Recently Christopher Fuchs and Asher Peres seem to argue in a related direction,
performing a nice
exercise in ``interpretation-bashing''
\cite{Fuchs-Peres}, as I would call it.
If I interpret them correctly,
they assure that quantum theory is an operationally testable formalism
concerned with predictions and frequency counts which are ultimately based
on clicks in detectors, nothing more.
Indeed, it is almost tautologic that the only relevant part of an
applicable formalism is the formalism itself.
(Just as any physical system obviously is a perfect copy of itself.)
Any emanation of a physical system can be codified into
discrete detectable clicks.
The statistical accumulation of such detector clicks
serves as an interface to the quantum probabilities
which can be computed according to algorithmic (``cookbook-like'')
rules, no more but no less.
This is ultimate testability,
and in terms of predictions,
quantum theory has been standing out wonderfully for almost a century now,
with not the faintest indication of failure even in the remotest area of
phenemenology.

Any variation of this criticism against interpretation could be heard around the world, year round
and from many prominent physicists.
It has become almost fashionable to discredit interpretation.
Already Sommerfeld warned his students not to get
into these issues,
and, as mentioned by John Clauser \cite{clauser-talkvie},
not long ago scientists working in that field
had to be very careful not to become discredited as ``quacks.''
Richard Feynman \cite[p. 129]{feynman-law}
once mentioned the  {\em ``perpetual torment that results
from [[the question]], `But how can it be like that?' which
is a reflection of uncontrolled but utterly vain desire to see
[[quantum mechanics]] in terms of an analogy with something familiar''} and advised
his audience, {\em
``Do not keep saying to yourself, if you can possibly avoid it,
`But how can it be like that?'
because you will get `down the drain', into a blind alley from which nobody has yet
escaped.''}
Only recently,  some of that thinking and experimenting, as  a spin-off,
turned into  the highly respected and very active research areas
of quantum cryptography and computing.

Indeed, almost no research field
has been plagued with more fruitless
``meta'' papers and ``voodoo'' thoughts than quantum mechanics,
many such treatises written by uninformed laymen with merely
superficial working practice of quantum mechanics.
Surely, when reading these treatises, one is often reminded
of the pressing  question of scholasticism
how many angels fit on the top of a needle.
Many of these ``explanations'' simply misinform the public,
thereby discrediting science.
Not to mention the more radical ``quacks,'' who however seem
to favor relativity theory over quantum mechanics for their
vain attempts to unfold what appears to them
the occult  mysteries of physics.
These people either cannot understand  or do not accept
the rational scientific practices and mathematics.

Let us get back to interpretation-bashing.
Of course, there exists a ``quick and dirty'' rebuttal
to the ``No-Interpretation'' interpretation of quantum mechanics:
that it is also an interpretation, after all!
Just as atheism is also a system of belief whose central element
is the nonexistence of God.
Nevertheless, the matter is too serious to leave it there.
Because, if taken earnestly,
the suggestion to ``not interpret quantum mechanics''
could severely hamper research in this field.

Abandoning interpretation issues from quantum physics proper
amounts to throwing the baby out with the bath water, and to censorship at worst.
Why? Because
we would be at a loss in extending the present knowledge base of
physical theory to new phenomenological domains!
Stated pointedly, abandonment of interpretation amounts to imposing
restrictions to what is considered ``legal'' science.
Those stop signs for speculative thinking could be detrimental,
crippling the mind of the geniuses whose precious
weird ideas are the root for scientific revolutions and progress.
So often in history the contemporary
peers  have committed themselves to what appeared to them
the consolidated canon of knowledge,
and too often such attempts have turned out to be vain at best and scientific
impediments at worst.
Indeed, I believe that the  danger of such impediments by restricting
physics proper is higher than
the annoyance and distraction caused by useless interpretations.

Interpretation is to the formalism
what a
scaffolding in architecture and
building construction is to the completed building.
Very often the scaffolding has to be erected because
it is an indispensable part of the building process.
Once the completed building is in place, the scaffolding is torn down and
the opus stands in its own full glory.
No need for auxiliary scaffold any more.
But beware of those technicians who claim to be able to erect
some of those ``international style skyscrapers''
without any poles and planks!

I claim that ``soft'' interpretation issues very often serve as
a guideline or  intuition, which is absolutely necessary for
the quantum mechanical research program.
Almost by definition, a ``good'' interpretation
fosters research, while a ``bad'' one cripples it.
Thereby, I would definitely
not like to link interpretation to ``reality'' or ``truth,''
but would keep a very pragmatic attitude: as long as an interpretation,
as weird as it may sound, fosters research, it is a ``good'' one,
otherwise it may just be a waste of time. However,
as pointed out by Lakatos \cite{lakatosch} so often,
the contemporary peers may not be able to acknowledge
what sooner or later will become a progressive research program,
and one is left with the request for tolerance and an openness for new ideas.

Let me give some examples.
Often one hears that many researchers in
the quantum computation community
profit from the intuition obtained from the Everett interpretation.
Other researchers in quantum theory seem to favor the information aspects
of the processes involved, in particular the reversibility (one-to-oneness)
of the unitary time evolution.
By and large, my experience is that there exists as many interpretations
of quantum mechanics as there are physicists (the
author's own fancy can be read in \cite{svozil-2001-convention}).
Even those claiming to cling to the Copenhagen interpretation
have their own very original thoughts about and additions to it (e.g.,
consider Anton Zeilinger's
foundational principle for quantum mechanics \cite{zeil-99}).
Other examples are the pictorial representation of
the formalism, such as Feynman diagrams in the theory of quantized fields,
or interferometric schemes in quantum optics  \cite{green-horn-zei}.

I would like to suggest a more tolerant attitude towards interpretation
also in other areas of physical research,
in particular in relativity theory.
The ether interpretation, for example, nowadays appears to be highly
exotic, although it has been taken quite seriously by scientists such as
Paul Dirac \cite{dirac-aether}, John Bell \cite{bell-92} and
even Albert Einstein \cite{einstein-aether}, to name but a few.

Without interpretation and intuition,
the application of the formalism would be restricted to
automatic proofing, to an ``a thousand monkeys typing-away scenario''.
A similar hope was expressed around 1900 by the mathematical formalists,
most notably by David Hilbert, who expected to be able to find a final,
finite system of axioms
from which all true mathematical theorems could be deduced
by similar methods as recipes in a cookbook.
Today we would call such methods ``algorithmic.''
It might be a tedious task, but eventually every
mathematical theorem would be uncovered automatically; even Andrew Wiles' proof
of Fermat's last theorem, and also the theorems by Gleason, Bell and Kochen \& Specker,
to name but a few.
Eventually, G\"odel, Tarski, Turing, Chaitin and others have put an end to
the formalist's hope for an ultimate, all-encompassing theory of
everything in mathematics.
Their findings
have given way to more realistic possibilities of constructive proofing.

Suppose you could program a computer to handle complicated quantum computations
by a very general knowledge base of quantum theory.
Would you really believe that any such computer
would become a good quantum theorist?
My suspicion is that this machine would churn out zillions
of tautologies and correct but unusable expressions.
By the time it would start producing important revelations (if ever),
the machine would have developed its intrinsic ``meaning''
or ``understanding,'' its ``interpretation'' of quantum mechanics.
This might be considered as a third, independent type of test for intelligence,
besides the famous Turing test or Daniel Greenberger's test of intelligence
(\cite{greenberger-testint};
in short: disobedience  of superrules, just as in the Genesis).

Of course, by mere brute force, eventually, every automatic, algorithmic
method would result in all possible theoretical expressions.
But then, who could filter out the scarce singular important treasures
from the myriads of  formulae,
the promising technological and experimental applications from the wasteful
repetitions of useless by-products?
At least until now, the quantum formalism
does not contain any such meta-formalisms.
Presently, this work is done by humans
with intuition and interpretations in their minds.
I cannot see any presently existing
(meta-)theory or machine which would be capable of taking away that selection
challenge and burden from the scientist.
There is so much creativity
required in applying scientific results into practical use
that I have no high hopes for any fast  automated solution.

Let me conclude with a partial answer to the question of the title,
``What could be more practical than a good interpretation?''
I would argue that proposals of a new theory
which can be distinguished
from quantum mechanics and extends it  experimentally would be more exciting
than any new interpretation.
Likewise,
any unexplained phenomenon which gives a hint to think about
new directions would be most welcomed.
At the moment, I am at a loss of seeing any one of these possibilities
(but maybe it is just my ignorance which prevents me from doing this).
In the meantime, let us be tolerant to all those weird interpretations of
quantum mechanics out there, since they are the guarantee for a
vivid and abundant scientific research.
In doing this, one should not accept claims of absolute truth of any particular
interpretation for two reasons:
no interpretation can be distinguished from
other interpretations by operationalizable means,
and it may be just a scaffolding, after all!
I readily acknowledge that any trickery quantum talk may degenerate to
fruitless speculations.
Yet, in view of the mindboggling \cite{green-horn-zei},
incomprehensiveness  \cite[p. 129]{feynman-law} of quantum mechanics,
any outright negation of interpretation may amount to a mistake mentioned already
in Democritus of Abdera's fragments
(translated by Cyril Bailey \cite{bailey-cyril}
cited by Erwin Schr\"odinger \cite[p. 87]{schroed:natgr}),
in which the senses, when attacked by reason, say,
{\em ``$\cdots$
from us you are taking the evidence by which you would overthrow us?
Your victory is your fall.''}

%\bibliography{svozil}
%\bibliographystyle{apsrev}

%Barnes, J. Early Greek Philosophy (Harmondsworth: Penguin), 1987.

%\begin{thebibliography}{99}
%\bibitem{WooFie} W.~K.~Wootters and B.~D.~Fields: \emph{Optimal State-Determination by Mutually Unbiased Measurements}, Ann.Phys. 191,363-381 (1989)
%\end{thebibliography}
\end{document}